# One laser pulse generates two photoacoustic signals

Fei Gao[1,†,*], Xiaohua Feng[1,†], Linyi Bai[2], Ruochong Zhang[1], Siyu Liu[1], Ran Ding[1], Rahul Kishor[1], Yanli Zhao[2,3], Yuanjin Zheng[1,*]

[1]School of Electrical and Electronic Engineering, Nanyang Technological University
[2]School of Physical and Mathematical Sciences, Nanyang Technological University
[3]School of Materials Science and Engineering, Nanyang Technological University

[†]Authors contributed equally to this work

[*]Corresponding to: fgao1@e.ntu.edu.sg and yjzheng@ntu.edu.sg

## Abstract:

Photoacoustic sensing and imaging techniques have been studied widely to explore optical absorption contrast based on nanosecond laser illumination. In this paper, we report a long laser pulse induced dual photoacoustic (LDPA) nonlinear effect, which originates from unsatisfied stress and thermal confinements. Being different from conventional short laser pulse illumination, the proposed method utilizes a long square-profile laser pulse to induce dual photoacoustic signals. Without satisfying the stress confinement, the dual photoacoustic signals are generated following the positive and negative edges of the long laser pulse. More interestingly, the first expansion-induced photoacoustic signal exhibits positive waveform due to the initial sharp rising of temperature. On the contrary, the second contraction-induced photoacoustic signal exhibits exactly negative waveform due to the falling of temperature, as well as pulse-width-dependent signal amplitude which is caused by the concurrent heat accumulation and thermal diffusion during the long laser illumination. An analytical model is derived to describe the generation of the dual photoacoustic pulses, incorporating Gruneisen saturation and thermal diffusion effect, which is experimentally proved. Lastly, an alternate of LDPA technique using quasi-CW laser excitation is also introduced and demonstrated for both super-contrast *in vitro* and *in vivo* imaging. Compared with existing nonlinear PA techniques, the proposed LDPA nonlinear effect could enable a much broader range of potential applications.

Photoacoustic (PA) technique has been attracting wide range of research interest in recent decade for biomedical imaging. Due to the hybrid merit of optical absorption and ultrasound detection, PA technique has successfully overcome the long-standing challenge of optical diffusion in deep scattering medium, achieving sensitive optical absorption contrast and scalable spatial resolution [1-7]. To induce strong enough PA signal, high-power nanosecond pulsed laser is usually utilized as the light source [8-11]. With ultrashort laser pulse (e.g. 1~10 ns), both thermal and stress confinements are satisfied to obtain optimized conversion efficiency from light absorption to ultrasound emission. When one pulse is illuminated on the target, one PA signal could be induced through transient absorption, heating and thermal expansion. However, there is no literature reporting using one laser pulse to generate two nonlinearly correlated PA signals. We report a PA technique based on long laser pulse induced dual PA (LDPA) nonlinear effect, which, for the first time, enables two nonlinearly correlated PA signals' generation from one laser illumination. To induce two PA signals from one laser pulse, two sharp laser intensity transitions are required. Therefore we employ square-profile laser illumination, rather than Gaussian-profile of conventionally used laser source. Due to the sharp rising and falling edges of the square laser pulse, two PA signals originating from thermal expansion and contraction respectively, are expected to be generated. In addition, the laser pulse width should be long enough (larger than stress confinement time), so that the expansion-induced and contraction-induced PA signals could be separated in time domain. It is also expected that the two PA signals should show phase-inverted (180 degree difference) waveforms due to the separate expansion and contraction effects. More interestingly, the two PA signals should exhibit different signal strengths that may vary with different laser pulse width, which are caused by the concurrent heat accumulation and diffusion during the long laser pulse illumination (thermal confinement is not strictly met). Being clearly distinct from other reported thermal nonlinear PA technique that only happens when laser fluence exceeds a threshold value [12-14], the proposed LDPA nonlinear effect could be observed with low-power laser diode and applied in much broader range of applications. The nonlinearity of the LDPA effect could immediately enables several interesting applications, such as axial super-resolution PA imaging [15], and PA-guided optical focusing [16]. Moreover, the pulse-width-dependent feature of the LDPA nonlinear effect could provide a characterization tool of different materials, and more importantly, it could also give design guideline for synthesizing advanced materials and nanoparticles to enhance the PA nonlinearity. In this paper, an analytical model is derived to describe the proposed LDPA nonlinear effect, and proof-of-concept experiments are performed to demonstrate its feasibility. The potential applications utilizing the LDPA effect are also discussed.

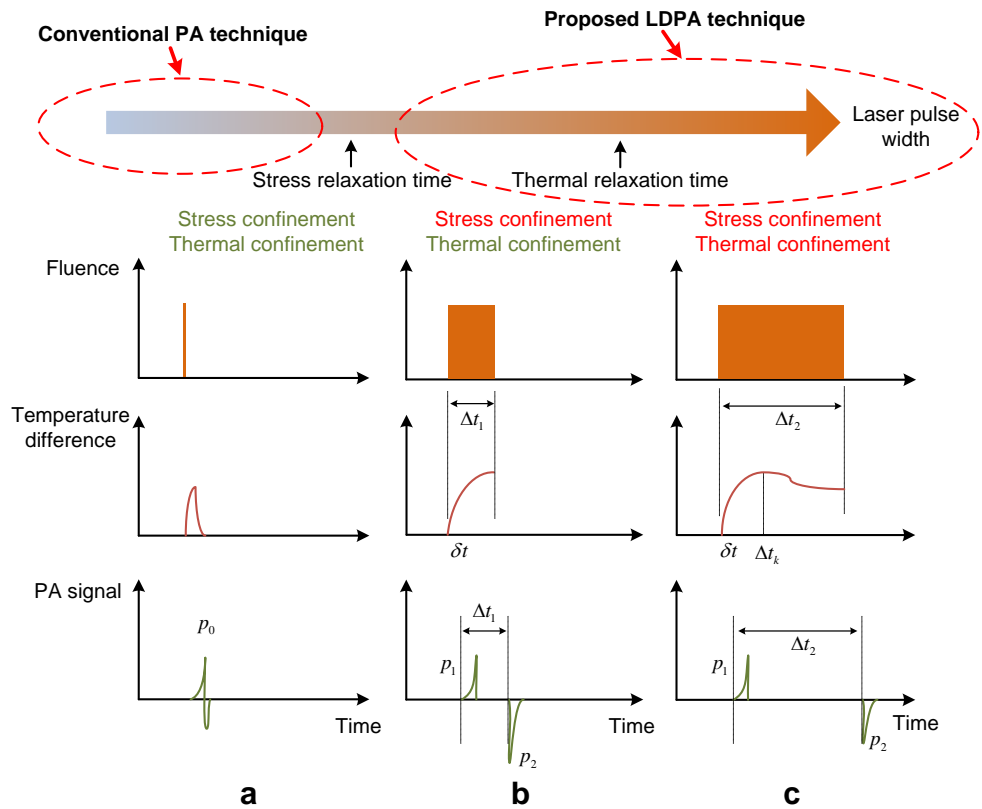

**Figure 1** The fluence pattern, temperature change and PA signal waveform of (a) conventional short laser pulse induced PA effect with both stress and thermal confinements satisfied; (b) long laser pulse induced dual PA nonlinear effect with only thermal confinement satisfied, and (c) even longer laser pulse without thermal confinement.

## Results

Conventionally the short laser pulse with nanosecond pulse-width is used to induce transient temperature change in the optical absorber with both stress and thermal confinements, and a dipolar PA signal is generated following the thermoelastic expansion and contraction (**Fig. 1a**). To further increase the laser pulse-width to be larger than stress relaxation time ($\Delta t_1 > \tau_s = d/v$, $d$ is the PA source diameter, $v$ is the acoustic velocity), the single dipolar PA signal is gradually separated into two independent but correlated PA signals. The first pulse is expansion-induced positive PA signal $p_1$, the second pulse is contraction-induced negative PA signal $p_2$ with exactly inverted waveform (**Fig. 1b**). The first PA signal is generated immediately after the sharp temperature rise of the laser illumination, which is related with the light irradiation $\phi$ (W/cm$^2$) within short rising period $\delta t$. The analytical expression of $p_1$ is similar with the conventional short laser pulse induced PA:

$$p_1 = \Gamma_0 \eta_{th} \mu_a \phi \delta t, \qquad (1)$$

where $\Gamma_0$ is the Gruneisen coefficient at the ambient temperature. $\eta_{th}$ is the conversion efficiency from heat to pressure, and $\mu_a$ is the optical absorption coefficient.

**ArXiv preprint version**On the other hand, the second PA signal $p_2$ could exhibit larger amplitude than the first one, which could be caused by more light energy deposition in the object, and higher Gruneisen coefficient (thermal expansion coefficient increased) due to the temperature rising of the object (**Fig. 1b**). However, when the laser pulse-width is further increased to larger than thermal relaxation time ($\Delta t_2 > \tau_{th} = d^2/\alpha$, $\alpha$ is the thermal diffusivity of the object), the amplitude of $p_2$ will be close to the amplitude of $p_1$, which is due to the significant thermal diffusion during the long laser illumination. Incorporating the Gruneisen increase, heat deposition and diffusion, the analytical expression of $p_2$ could be derived as below [15, 16]:

$$p_2 = \left\{\Gamma_0 + \overbrace{b\eta_{th}\mu_a\tau_{th}^2\phi\left[1-\left(1+\frac{\Delta t}{\tau_{th}}\right)e^{-\frac{\Delta t}{\tau_{th}}}\right]}^{\textit{Gruneisen saturation term}}\right\}\eta_{th}\mu_a\phi\delta t + \overset{\textit{Fluence}}{\Gamma_0\eta_{th}\mu_a\phi\Delta t}\ \overset{\textit{Thermal diffusion}}{e^{-\frac{\Delta t}{\tau_{th}}}}, \qquad (2)$$

where $b$ is the coefficient relating the absorbed thermal energy to the Gruneisen parameter change, and $\Delta t$ is the laser pulse-width. Compared with Eq. (1) of the first conventional linear PA, the second PA signal $p_2$ shows nonlinearity governed by three terms in Eq. (2). The Gruneisen saturation term describes the object's temperature rising and saturation due to heat accumulation during one laser pulse illumination. The light fluence term is linearly increasing with laser pulse-width, and the thermal diffusion term is exponentially decaying according to Newton's law of cooling due to temperature difference. The Gruneisen saturation term is derived by the time integration of heat deposition and diffusion (**Supplementary Note 1**). A quick observation of Eq. (2) shows that with increasing pulse width $\Delta t$, the first term with Gruneisen saturation in Eq. (2) increase. However, the second term in Eq. (2) may increase first due to the linearly increasing fluence, and then exponentially decrease due to the thermal diffusion. To analyze Eq. (2) quantitatively, we simplify the Eq. (2) by substituting some parameters with three constants: $A = \Gamma_0 \eta_{th} \mu_a \phi \delta t = p_1$, $B = b\eta_{th}^2 \mu_a^2 \phi^2 \delta t$, $C = \Gamma_0 \eta_{th} \mu_a \phi$. Then Eq. (2) could be expressed as:

$$p_2 = p_1 + B\tau_{th}^2 \left[1 - \left(1 + \frac{\Delta t}{\tau_{th}}\right) e^{-\frac{\Delta t}{\tau_{th}}}\right] + C\Delta t e^{-\frac{\Delta t}{\tau_{th}}} \tag{3}$$

To obtain the optimum laser pulse width for maximizing the amplitude of $p_2$, the derivative of Eq. (3) is calculated (**Supplementary Note 2**) and expressed as below:

$$\frac{\partial p_2}{\partial \Delta t} = \left[C + \left(B - \frac{C}{\tau_{th}}\right)\Delta t\right] e^{-\frac{\Delta t}{\tau_{th}}} \tag{4}$$

To let $\partial p_2 / \partial \Delta t = 0$ for maximum $p_2$ amplitude, we obtain the optimum laser pulse width:

$$\Delta t_0 = \frac{C}{\frac{C}{\tau_{th}} - B} \tag{5}$$

It indicates that the optimum laser pulse width $\Delta t_0$ doesn't always have a positive value. To be clear, the analysis is divided as below cases:

1) $C < B\tau_{th} \Rightarrow \Delta t_0 < 0$

For this case, there is no positive optimum laser pulse width. From Eq. (3) and smaller constant $C$ than $B\tau_{th}$, the Gruneisen saturation term will dominate, and Eq. (3) could be simplified as:

$$p_2 = p_1 + B\tau_{th}^2 \left[1 - \left(1 + \frac{\Delta t}{\tau_{th}}\right)e^{-\frac{\Delta t}{\tau_{th}}}\right] \tag{6}$$

It shows that with increasing laser pulse width, the amplitude of the second LDPA signal is expected to increase accordingly until saturation (**Fig. 1b**). This case is similar with the thermal nonlinear PA technique using two consequent short high-power laser pulses, where the second PA pulse is nonlinearly enhanced by the Gruneisen relaxation effect induced by the first pulse's heating [15, 16]. However, the proposed method could enable significant thermal nonlinearity by simply tuning the laser pulse width for sufficient heating accumulation using low power CW laser, and the temperature can easily rise sufficiently for nonlinear PA imaging, even reach tens of degrees increase for concurrent photothermal therapy. On the other hand, the two short consequent pulse method requires ultrahigh laser pulse energy and tight focusing to induce obvious temperature rise for the thermal nonlinear phenomenon.

Simulation under the condition $C < B\tau_{th}$ was conducted by sweeping the laser pulse width with different thermal relaxation time (20~100 μs) to calculate the nonlinear PA incremental amplitude ($p_2 - p_1$). The result shows that with increasing laser pulse width, the nonlinear PA amplitude increases gradually towards saturation (**Fig. 2a**), which is due to the heat accumulation and temperature rising. For larger value of thermal relaxation time, because of better thermal confinement property and heat accumulation, the nonlinear PA incremental amplitude increases more significantly than that with smaller thermal relaxation time. On the other hand, if the local temperature exceeds the boiling threshold of the medium (e.g. water) before saturation, bubble will be generated thereafter. It is worth noting that although longer laser pulse width is preferred to maximize the LDPA signal, the amplitude will reach saturation and there will be no significant increase when further increasing pulse width. As indicated by the dotted arrow line (**Fig. 2a**), A larger value of thermal relaxation time exhibits a longer increasing period, which requires longer laser pulse width to reach the saturation.

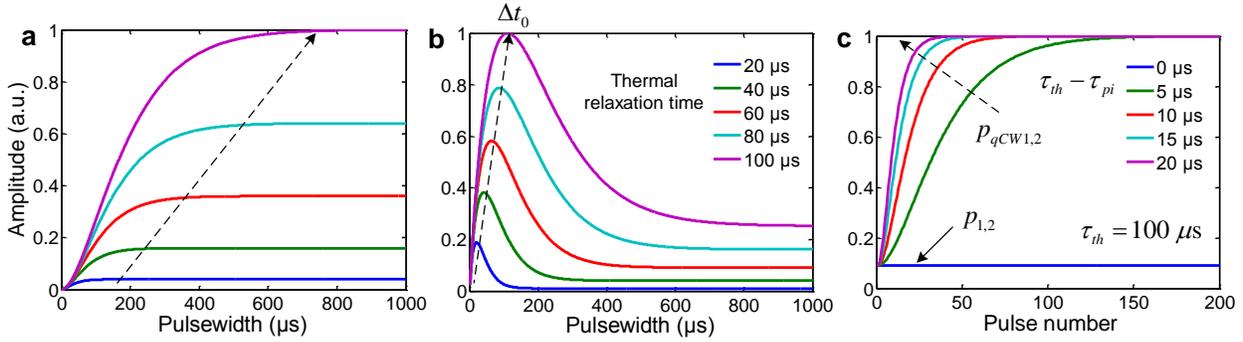

**Figure 2** (a-b) The amplitude of the nonlinear PA incremental amplitude versus laser pulse width with different thermal relaxation time under case 1 and case 2. The dotted arrow line indicates the laser pulse width when the maximum LDPA amplitude is achieved. (c) The amplitude of the nonlinear PA amplitude versus laser pulse number with different pulse-pulse intervals (thus different $\tau_{th} - \tau_{pi}$) under case 3 with constant thermal relaxation time 100 μs.

2) $C > B\tau_{th} \Rightarrow \Delta t_0 > 0$

For this case, there will be a positive optimum laser pulse width $\Delta t_0$ for maximum LDPA signal. When $C \gg B\tau_{th}$, the Eq. (3) and Eq. (5) could be simplified as $p_2 = C\Delta t e^{-\frac{\Delta t}{\tau_{th}}}$ and $\Delta t_0 \approx \tau_{th}$, where $\Delta t_0$ is the turning point from amplitude increase to decrease of the LDPA signal (**Fig. 1c**), which could be used as the nonlinear characterization feature. The straightforward way to characterize this PA nonlinearity is to sweep the pulse width $\Delta t$ and analyze the amplitude of $p_2$. As expected, simulation result shows that with increase laser pulse width, the amplitude experiences increasing and then decreasing accordingly (**Fig. 2b**). Moreover, with larger thermal relaxation time indicated by the dotted arrow line (**Fig. 2b**), the nonlinear PA incremental amplitude increases significantly due to better thermal confinement capability of the target. Meanwhile, the key pulse width $\Delta t_0$ also increases for maximum PA nonlinearity.

3) Higher pulse repetition rate in quasi-CW mode

Except the heat accumulation and diffusion within one laser pulse, the long-term heat accumulation and base temperature rising could also be significant when the laser pulse repetition rate (N pulses per second with pulse-pulse interval $\tau_{pi} = 1/N$) is increased. Both $p_1$ and $p_2$ will be affected by another quasi-CW temperature rise and saturation term as a function of pulse number $p_2$ and could be expressed as:

$$p_{qCW1,2} = p_{1,2} + B\tau_{th}^2 \left\{ 1 - \left[ 1 + \frac{(n-1)(\tau_{th} - \tau_{pi})}{\tau_{th}} \right] e^{-\frac{(n-1)(\tau_{th} - \tau_{pi})}{\tau_{th}}} \right\}, \tau_{th} - \tau_{pi} > 0 \tag{7}$$

It deserves noting that $\tau_{th} - \tau_{pi} > 0$, so that within two consequent laser pulses, the accumulated heat by the first laser pulse doesn't totally dissipate before arrival of second laser pulse. When $\tau_{th} - \tau_{pi} = 0$ or $n = 1$, Eq. (7) goes back to $p_{qCW1,2} = p_{1,2}$ due to no heat accumulation between consecutive pulses or only single pulse. Eq. (7) shows that the absolute temperature change caused by the increased number of laser pulses in quasi-CW mode will nonlinearly enhance the signal amplitude of both $p_1$ and $p_2$ until saturation at a fixed laser pulse width. This indicates the potential to immediately enable the concurrent temperature detection and close-loop control during photothermal treatment [17, 18]. In addition, it also shows that even for the linear PA signal $p_1$ with increased laser repetition rate working in a quasi-CW way, $p_{qCW1}$ could also exhibit nonlinearity. The simulation results with constant thermal relaxation time 100 μs and different pulse-pulse intervals (0~20 μs) show that when increasing number of laser pulses, the amplitude of $p_{qCW1,2}$ increases until saturation (**Fig. 2c**). As expected, when $\tau_{th} - \tau_{pi} = 0$, the amplitude keeps constant due to no heating accumulation within pulse-pulse intervals. Lastly, as indicated by the dotted arrow line with $\tau_{th} - \tau_{pi}$ increasing, less number of laser pulses is required for the nonlinear PA signal to reach the maximum amplitude.

The experimental setup is like a typical PA microscopy system (**Fig. 3a** and **Methods**). The first observation of the LDPA nonlinear effect is from a black rubber wire sample, which has strong optical absorption at the laser's wavelength. For comparison, the conventional single PA signal could be generated by short laser pulse width. With 1 μs laser pulse width (**Fig. 3b**), only one PA signal could be generated (**Fig. 3d**). On the other hand, a typical measured waveform of LDPA signal with 10 μs laser pulse width (**Fig. 3c**) is shown in **Fig. 3e**. As expected, the delay between the two PA signals generated from rising and falling edges of pulse laser illumination is also 10 μs. The waveforms of the two PA signals show high correlation and inverted amplitude, which is caused by the initial expansion (first PA signal) and contraction (second PA signal) from the same optical absorber. In addition, the peak-to-peak amplitude of the second nonlinear PA signal is larger than that of first linear PA signal, which results from the heating accumulation and temperature rising during the laser pulse.

By sweeping the laser pulse width from 6 μs to 900 μs, the amplitudes of both first and second PA signals from the black rubber wire are recorded (**Fig. 3f**). As expected, the first PA signal's amplitude keeps almost constant when sweeping the laser pulse width, because the expansion-induced PA signal is generated by the rising edge of light illumination, and not influenced by the heat accumulation and diffusion during the long-pulse light illumination. However, the second PA signal, as well as the ratio between the two PA signals, experiences an increase (<10 μs), flat (10~20 μs) and decrease (20~500 μs) when sweeping the laser pulse width. The measurement result agrees well with the prediction in **Fig. 2b** when the thermal relaxation time of the medium is small ($C > B\tau_{th}$).

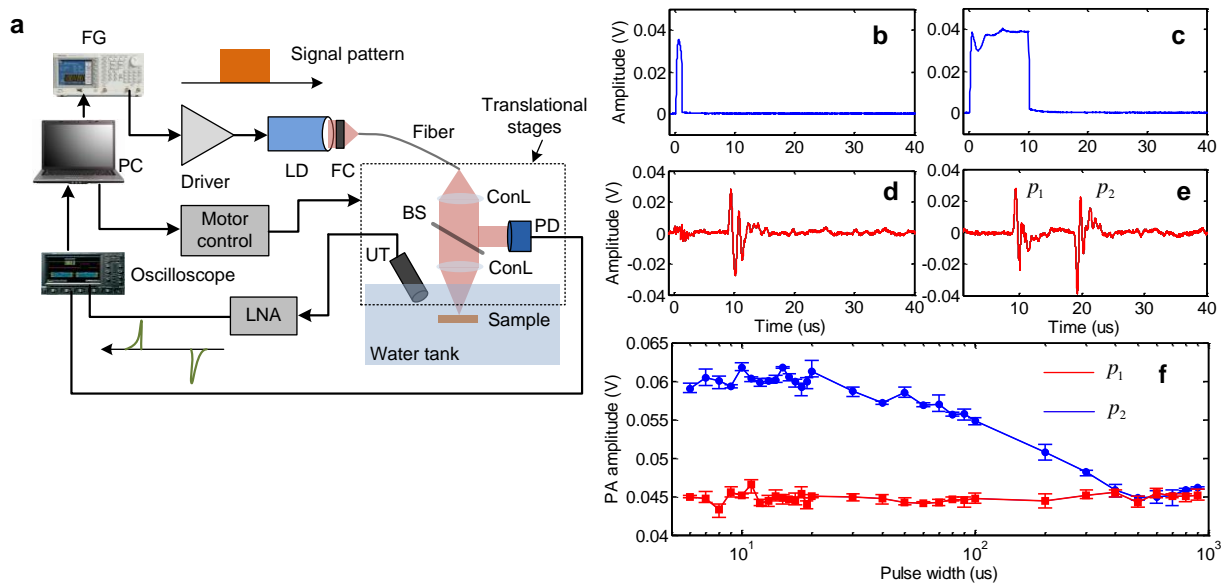

**Figure 3** (a) The schematic of the experimental setup. FC: function generator; LD: laser diode; FC: fiber coupler; ConL: condenser lens; BS: beam splitter; PD: Photodiode UT: ultrasound transducer; LNA: low-noise amplifier. (b)-(c) The light illumination signal detected by the photodiode with pulse width of 1 μs and 10 μs. (d)-(e) The measured LDPA signals with 1 μs and 10 μs pulse width laser illumination. (f) Amplitudes of first PA signal (red line, square symbol) and second PA signal (blue line, circle symbol) by sweeping the pulse width of the laser source.

As predicted from **Fig. 2a**, when the thermal relaxation time of the medium is large enough ($C < B\tau_{th}$), the second nonlinear PA signal will keep rising when increasing the laser pulse width. To demonstrate this situation, we took black ink sealed in a plastic micro-tube as the testing sample to be immersed in water. Due to better heat isolation of the plastic layer compared with the previous black wire sample contacting water directly, the sealed black ink sample is expected to achieve much larger thermal relaxation time. By tuning the laser pulse width to be 100 μs, 500 μs and 1 ms, it is observed that the second nonlinear PA signal keeps increasing with larger laser pulse width as expected (**Fig. 4a-c**). Finally, when laser pulse width was further tuned to be 1.5 ms, the local temperature at the focal point reached the boiling point of water leading to bubble generation (**Fig. 4d**) and significantly enhanced PA signal generation. The comparison plot shows that the nonlinear ratio ($p_2/p_1$) of bubble-induced PA is around 450, much greater than the LDPA signals before bubble generation (**Fig. 4e**). The upper limit of the nonlinear ratio for the LDPA signals generated from water is around 3.5~4, which equivalently refers to the temperature rise from ambient temperature (~25 °C) to boiling point (~100 °C). To explore the temperature-dependent property of the LDPA effect, the repetition rate of the laser source (fixed pulse width: 10 μs, peak power: 1 W) was swept from 100 Hz to 1.9 kHz, which corresponds to average power sweeping from 1 mW to 19 mW. **Fig. 4f** shows the amplitudes of the two PA signals versus average power sweeping. As expected, with increasing average power of the laser source, absolute temperature of the sample also increases. Then the elevated Gruneisen coefficient enhances both PA signals, which is predicted from **Fig. 2c**.

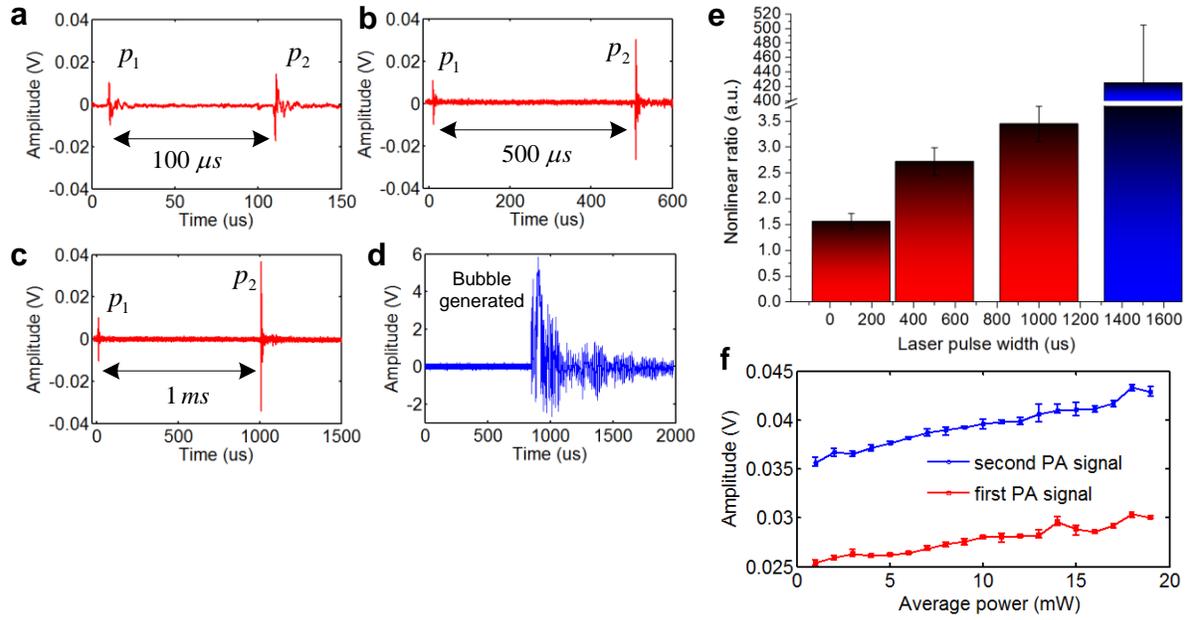

**Figure 4** The measured LDPA signals from black ink with laser pulse width of (a) 100 μs, (b) 500 μs and (c) 1 ms. (d) Significant PA signal generation due to bubble generation and collapse with 1.5 ms laser pulse width. (e) The nonlinear ratio comparison with different laser pulse widths. (f) Amplitudes of first PA signal (red line, square symbol) and second PA signal (blue line, circle symbol) by sweeping the average power of the laser source.

Although the heat accumulation and nonlinear PA enhancement could be obviously observed in the above measurement, the nonlinear PA signal's amplitude is relatively weak due to the low peak power of the CW laser source. To compensate the PA signal's amplitude as well as utilizing the merit of the proposed LDPA effect, here we further propose an alternative LDPA technique by using quasi-CW laser – a high-repetition-rate pulsed laser diode. As shown in **Fig. 5a**, the laser source excites $N$ consecutive short laser pulses with high repetition rate, and the time interval between two adjacent laser pulses are short enough for continuous heat accumulation (the heat induced by previous laser pulse doesn't totally dissipate before the next laser pulse reaches). Therefore, the local temperature rising, as well as the thermal expansion coefficient of the object, could continuously accumulate during the N laser pulse sequence, leading to nonlinear enhanced PA signal amplitude.

By replacing the CW laser in the previous setup with a quasi-CW laser source (**Methods**), 500 consecutive laser pulses were illuminated on the black ink sample. As shown in **Fig. 5b**, the amplitudes of the 500 PA signals are increasing as expected, where the last nonlinear PA signal (**Fig. 5d**) exhibits significantly enhanced amplitude compared with the first linear PA signal (**Fig. 5c**). By quantitatively plotting the nonlinear increase percentage versus number of pulses in **Fig. 5e**, we can see that ~ 80% enhancement is achieved by the LDPA technique with quasi-CW laser source. In comparison, only ~ 3% enhancement is achieved by using two consecutive laser pulses in literature [15, 16]. It is expected that by further increasing the number of laser pulse, the nonlinear enhancement will increase and finally reach constant, which is caused by the balanced thermal accumulation and diffusion. Finally, imaging results were acquired by raster-scanning over a curved black wire sample based on the LDPA technique by quasi-CW laser. The reconstructed images by the first linear PA signal, the last nonlinear PA signal, and the differential image are shown in **Fig. 5f**. As indicated in **Fig. 5g**, the image contrast of the nonlinear PA imaging shows a slight improvement (6.7:1) than the linear PA imaging (5.9:1). However, the differential image contrast achieves a much greater improvement (14.8:1). The underlying reason is that by subtracting the linear PA image from nonlinear PA image, the pure nonlinear increase caused by the heat accumulation is extracted. At the same time, the linear absorption background is significantly suppressed by the subtraction [19-21].

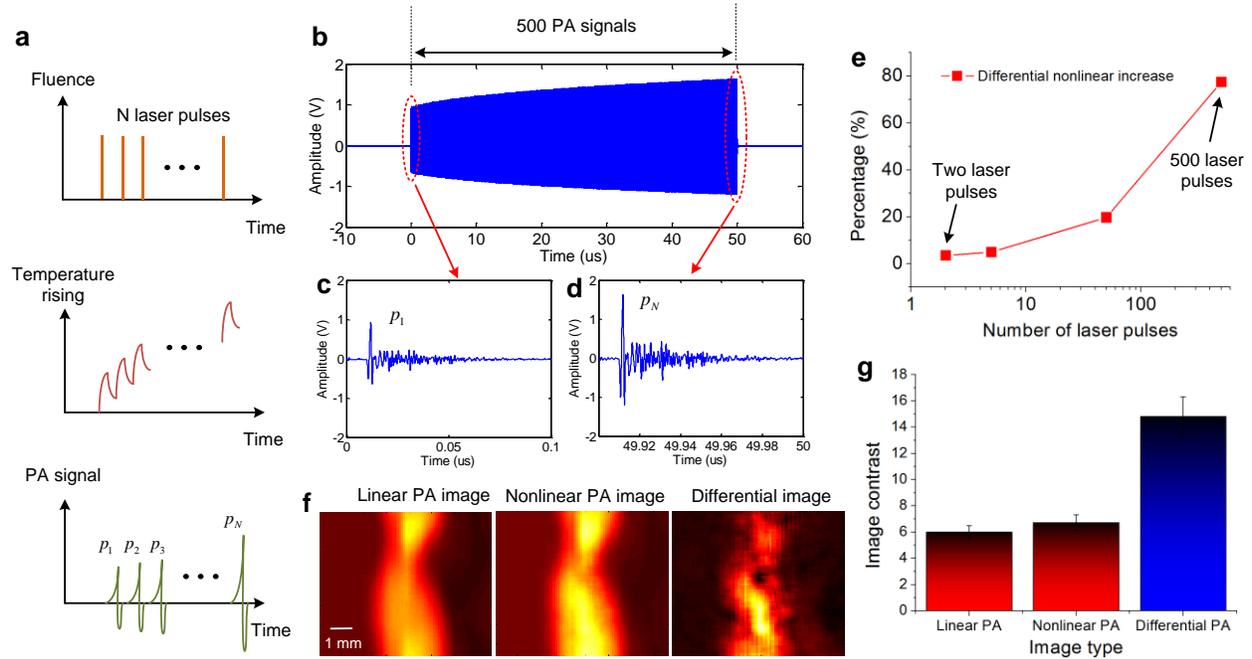

**Figure 5** (a) The fluence pattern, temperature rising and PA signal diagram of the proposed LDPA technique by quasi-CW laser source. (b) A typical 500 PA signals, (c) the first linear PA signal, and (d) the last nonlinear PA signal. (e) The percentage of nonlinear increase versus the number of consecutive laser pulses. (f) The linear PA imaging, nonlinear PA imaging and differential imaging of a curved black wire, (g) and their quantitative comparison of imaging contrast.

Finally, *in vivo* imaging in rat based on the proposed LDPA nonlinear effect was performed to demonstrate the feasibility towards biomedical imaging applications. We developed the PA imaging system based on raster scanning and quasi-CW laser excitation (**Fig. 6a** and **Methods**) to acquire linear PA imaging, nonlinear PA imaging and differential imaging in single modality. Seven-week-old rat was used for the in vivo experiments with subcutaneous injection of a kind of NIR organic dye – IR-820 into the abdomen region (**Supplementary Fig. S1** and **Methods**). As shown in **Fig. 6b**, due to the existence of blood spot at the surface of the abdomen skin, very strong PA signal was detected in the linear PA imaging result. However, the linear PA signal from the IR820 is overwhelmed by the strong PA signal from blood and not discernable. Then, the nonlinear PA image was acquired by reconstructing the last nonlinear PA signal of the pulse sequence. It is observed that the nonlinear PA signal from IR820 increases significantly due to its better thermal confinement and heat accumulation under the skin. On the other hand, the nonlinear PA signal from the blood is still very strong, but has subtle increase over their linear PA signals. By applying differential imaging, the image contrast of the IR820 over the blood background is further improved by suppressing the PA signal from blood with weak nonlinearity (**Fig. 6c**). The quantitative comparison shows that the image contrast is improved from 0.49 (linear PA imaging) to 1.7 (nonlinear PA imaging) and 12.3 (differential imaging) (**Fig. 6d**).

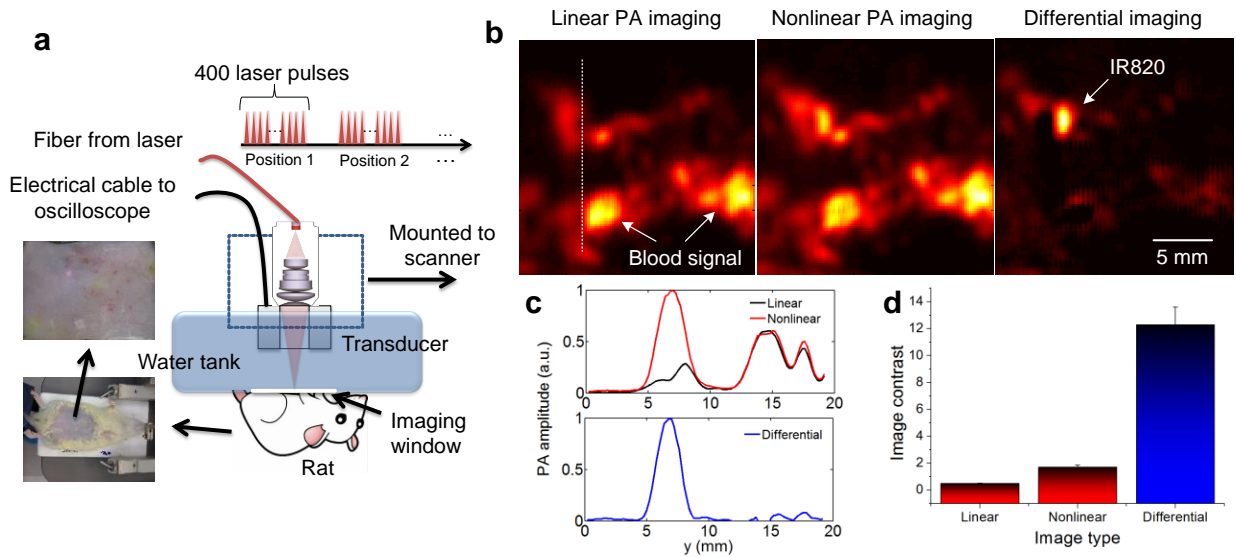

**Figure 6** (a) in vivo experimental setup based on the LDPA technique by quasi-CW laser excitation. (b) the reconstructed linear PA, nonlinear PA and differential imaging results of a rat with subcutaneous injection of IR820. Strong PA signals from high-absorptive blood background are shown in the linear and nonlinear PA images. By subtracting, the signal from the IR820 pops up with linear background suppression. (c) Normalized signal intensity of the linear, nonlinear and differential images along the white dashed line in b. (d) The image contrast comparison of the three images in b.

## Discussion

To better illustrate the merits of the proposed LDPA technique by both CW and quasi-CW laser source, a table is provided below to compare it with the recently proposed dual-pulse nonlinear PA technique [15, 16]:

Table I The comparison table between the proposed LDPA techniques with the existing dual-pulse nonlinear PA technique

|  | **Dual-pulse nonlinear PA** | **Proposed LDPA by CW laser excitation** | **Proposed LDPA by Quasi-CW laser excitation** |
|---|---|---|---|
| **Physical mechanism** | | | |
| **Nonlinear origin** | Impulse temperature rise then heat relaxation during laser interval | Continuous temperature rise and heat accumulation | Quasi-continuous temperature rise and heat accumulation |
| **Physical conditions** | Thermal and stress confinements satisfied for two PA generations | First PA signal: satisfied Second PA signal: unsatisfied | Thermal and stress confinements satisfied for all PA signals |
| **PA generation mechanism** | Two laser pulse induce two PA signals | One laser pulse induce two PA signals | $N$ consecutive laser pulses induce $N$ PA signals |
| **System performance** | | | |
| **Heating efficiency** | ×Low (single short laser pulse heating) | √High (continuous laser heating) | √High (Quasi-continuous laser heating) |
| **PA signal strength** | √High (high peak power pulsed laser) | ×Low (low-power CW laser) | √High (high peak power pulsed laser) |
| **System cost** | ×Very high (two high-power laser systems required) | √Very low (single CW laser diode) | √Low (single high-rep-rate pulsed laser diode) |
| **Applications** | ×Need tight focusing to improve the heating efficiency and nonlinearity (e.g. time-reversal optical focusing in scattering medium) | ×Need strong absorber and focusing to improve the PA SNR (e.g. contrast-enhanced PA microscopy) | √Generally applicable to most PA embodiments (e.g. PA tomography, microscopy, endoscopy, etc.) |

It shows that the three techniques differ in terms of both physical mechanism and system performance. Especially regarding heating efficiency and system cost, the proposed LDPA technique by CW laser exhibits much more advantageous than the existing dual-pulse nonlinear PA method. To overcome the low peak power of the CW laser source, the alternative LDPA technique by quasi-CW laser bridges the merits of the other two and achieves simultaneous high heating efficiency, high PA signal strength and low system cost. Therefore, the potential applications of the proposed LDPA effect could include, and go beyond the existing applications of dual-pulse nonlinear PA technique:

*LDPA effect for nonlinear PA tomography, microscopy, endoscopy, etc.:* With sufficient heating efficiency and PA signal strength, the LDPA technique by quasi-CW laser excitation is generally applicable to all existing linear PA imaging embodiments, pushing them towards nonlinear PA imaging with enhanced contrast and resolution.

*LDPA effect for nanoparticle-enhanced PA imaging:* Compared with conventional contrast-enhanced PA imaging utilizing nanoparticles with strong optical absorption, the nanoparticles with strong thermal confinement property is preferred to enhance the LDPA nonlinear effect. From the developed analytical model, the nanoparticles should show good thermal confinement (smaller thermal diffusivity), which may originates from single particle, and/or the conjugation of the nanoparticles to confine the laser induced heating [22].

*LDPA effect for close-loop photothermal treatment and photoacoustic temperature sensing:* As discussed, LDPA signal is generated from pulse-modulated CW or quasi-CW laser and is linearly proportional to the absolute temperature of the object. On the other hand, photothermal treatment is also usually performed by CW laser. Therefore, utilizing the LDPA effect, it could immediately enable close-loop photothermal treatment with PA temperature monitoring by self-regulating the repetition rate of the laser source [17].

*LDPA effect for axial-resolution improvement of PA imaging:* Based on the theory of Gruneisen relaxation PA microscopy [15], the proposed LDPA effect is also potential to improve axial-resolution of PA imaging. The subtraction of the two PA signals will render a narrower axial full width at half maximum (FWHM) due to its optical sectioning capability.

*LDPA effect for PA-guided optical focusing in scattering medium:* Using the LDPA effect for PA-guided optical focusing in scattering medium is feasible according to the reported literature [16]. Instead of utilizing two high-power pulsed lasers, comparable performance could be expected by utilizing single low-cost CW laser source based on the LDPA effect.

In conclusion, we report the LDPA nonlinear effect with one pulse illumination by modulated CW laser and two PA signals detection. Then we extend it to an alternative LDPA technique by quasi-CW laser source for stronger PA signal strength. An analytical model is derived to model the LDPA effect. Then both *in vitro* and *in vivo* sensing and imaging have been performed to demonstrate the feasibility of the technique towards biomedical imaging applications. The proposed LDPA technique enables several interesting applications based on low-power low-cost laser diode, which will be further explored in future work.

## Methods

***In vitro* PA measurement based on LDPA effect by CW laser excitation.** As shown in **Fig. 3a**, the input square-wave pulse with tunable pulse width was generated by a function generator, which was connected to a custom-designed current driver (maximum current: 5 A). The output of the driver was then fed to a laser diode (wavelength: 808 nm, power: 1 W) with fiber coupling. The output light from the fiber was then collimated and weakly focused by condenser lens on the sample with spot size of ~500 μm. Meanwhile, a beam splitter and photodiode were utilized to monitor the laser intensity variation. An ultrasound transducer (Doppler Inc.) with 1 MHz central frequency was placed close to the sample. Both of the transducer and sample were immersed in water for optimum optical and acoustic coupling. The dual PA signals were firstly amplified by a low-noise amplifier (5662, Olympus), then recorded by an oscilloscope with 100 MSPS sampling rate.

***In vitro/vivo*** **PA imaging based on LDPA effect by quasi-CW laser excitation.** The *in vivo* imaging setup (**Fig. 6a**) was updated based on the previous *in vitro* setup. Firstly, the CW laser source was replaced by a high-repetition-rate pulsed laser diode (Quantel Laser Diode Illuminator, wavelength: 808 nm, pulse width: 100 ns; rep. rate: 10 kHz; pulse interval: 100 μs; maximum pulse energy: 1 mJ) to output the consecutive laser pulse sequence working in a quasi-CW mode. Secondly, the imaging head was modified by integrating an objective lens and a custom-designed ultrasound transducer with a hole inside in a confocal way to maximize the sensitivity. Lastly, the water tank was modified to accommodate the needs for *in vivo* imaging. A hole was cut at the bottom of the water tank, and then sealed by a thin transparent polyethylene membrane, which was directly placed on top of the skin of the rat's abdomen. To form an image, the imaging head was mounted to a 2D mechanical scanner for raster scanning over the sample. The step size is 200 μm over an imaging area of 2 cm.

**Animal preparation.** A ten-weeks-old rat was used for this animal imaging study. This study conforms to the Guide for the Care and Use of Laboratory Animals published by the National Institutes of Health, USA and protocol approved by the Institutional Animal Care and Use Committee (IACUC), Nanyang Technological University. Through the experiment, the animal was anesthetized with 2% vaporized isofluorane. Hair around the region of abdomen was removed after hair removal lotion treatment. Then the IR-820 was injected into the rat by subcutaneous injection. After applying the ultrasound gel to optimize the acoustic coupling, the water tank was placed tightly on top of the rat's abdomen to ensure the good optic and acoustic coupling.

**IR-820 synthesis and characterization.** The IR-820 near-infrared dye and chloroauric acid (HAuCl4*3H2O) employed were commercially available and used without further purification unless specifically mentioned. IR-820 in aqueous solution exhibits peak absorption at 819 nm and still very strong absorption under the 808 nm wavelength of the laser used in the experiments.

**Author Contributions**